\documentclass[twoside,twocolumn,showpacs,preprintnumbers,aps]{revtex4}
\usepackage{graphicx}

\begin{document}

\title{Integral representations for nonperturbative generalized parton
distributions in terms of perturbative diagrams}
\author{P.~V.~Pobylitsa}
\affiliation{Institute for Theoretical Physics II, Ruhr University Bochum, D-44780
Bochum, Germany\\
and\\ Petersburg Nuclear Physics Institute, Gatchina, St.~Petersburg, 188300,
Russia}

\begin{abstract}
An integral representation is suggested for generalized parton distributions
which automatically satisfies the polynomiality and positivity constraints.
This representation has the form of an integral of perturbative triangle
diagrams over the masses of three propagators with an appropriate weight
depending on these masses. An arbitrary $D$ term can be added.
\end{abstract}

\maketitle

\section{Introduction}

Generalized parton distributions (GPDs)
\cite{MRGDH-94,Radyushkin-96-a,Radyushkin-96,Ji-97,Ji-97-b,CFS-97,Radyushkin-97,Radyushkin-review,GPV,BMK-2001}
play an important role in the QCD analysis of various hard phenomena such as
deeply virtual Compton scattering and hard exclusive meson production. GPDs
are defined in terms of nondiagonal hadron matrix elements of two quark
(gluon) fields separated by a light-like interval. GPDs contain a vast
amount of nonperturbative information about the quark-gluon structure of
hadrons. In particular, the usual forward parton distributions (FPDs) and
the form factors of hadrons can be expressed via GPDs.

In contrast with form factors and FPDs which can be directly accessed
experimentally, the case of GPDs is much more involved: the experimental
data can provide information only about some integrals containing GPDs. On
the theoretical side, GPDs are typical nonperturbative quantities. Although
there are no reliable methods for the calculation of GPDs from the first
principles of QCD, still the theory imposes certain constraints on GPDs
which should be taken into account in the analysis of the experimental data.
Among the general theoretical constraints on GPDs an important role is
played by the polynomiality of the Mellin moments \cite{Ji-97-b} and by the
positivity bounds \cite{Martin-98,
Radyushkin-99,PST-99,Ji-98,DFJK-00,Burkardt-01,Pobylitsa-01,Pobylitsa-02,Diehl-02,Burkardt-02-a,Burkardt-02-b,Pobylitsa-02-c,Pobylitsa-02-e}.

Since the GPDs cannot be measured directly, in the analysis of experimental
data one has to deal with models of GPDs. It would be preferable to use
those models which are compatible with the polynomiality and positivity
constraints. However, the realization of this idea meets with problems. The
polynomiality property holds automatically if one uses the so-called double
distribution representation for the GPDs
\cite{MRGDH-94,Radyushkin-96-a,Radyushkin-97}.
However, the double distribution
representation does not guarantee positivity. The general solution of
the positivity bounds is also known \cite{Pobylitsa-02-d} but the problem
remains how to describe the class of models of GPDs which satisfy both
positivity and polynomiality conditions.

In Ref.~\cite{Pobylitsa-02-d} an ansatz for GPDs is suggested which
automatically obeys both positivity and polynomiality constraints. The
method of Ref.~\cite{Pobylitsa-02-d} is based on a  formal mathematical
construction rather than on physical arguments. In this paper another approach
is taken.
We start from an analysis of simple perturbative graphs for GPDs. On general
grounds these graphs must obey both positivity and polynomiality
constraints. We check the positivity explicitly. Next we notice that the set
of functions obeying both polynomiality and positivity conditions is convex.
Therefore taking linear combinations of perturbative graphs for
\emph{different theories} weighted with positive coefficients we obtain new
solutions of the positivity and polynomiality constraints. The words
\emph{``different theories''} mean that we can average over various parameters:
masses, vertices, couplings, sets of fields, etc. At first sight this
approach looks like an artificial trick rather than physics. However, in this
paper we reveal certain structures standing behind the leading-order
perturbative graphs for GPDs in various theories and show that these
structures can be used as a sort of elementary blocks for the construction
of a rather wide class of models of GPDs obeying both polynomiality and
positivity constraints. The analysis is restricted to the case of spinless
hadrons (e.g. pions) but the methods suggested here allow a straightforward
generalization for the more interesting case of nucleon.

The structure of the paper is as follows.
Sec.~\ref{GPD-section} describes notations used for GPDs in the
usual and impact parameter representations.
Sec.~\ref{Poly-posi-section} contains a brief review of the positivity
and polynomiality properties of GPDs.
Sec.~\ref{general-method-section} explains how perturbative GPDs
can be used in order to construct solutions of the polynomiality and positivity
constraints on GPDs.
In Sections~\ref{triangle-phi-3} and \ref{triangle-Yukawa-section}
the leading order perturbative GPDs are analyzed in the $\phi^3$
and Yukawa models respectively. In Sec.~\ref{building-section}
integral representations are suggested for GPDs which automatically
obey the polynomiality and positivity constraints.
In Sec.~\ref{FPD-section} the consistency of the approach is tested
by checking the positivity of the corresponding forward parton
distributions.

\section{Generalized parton distributions}
\label{GPD-section}

The GPDs $H^{(N)}(x,\xi,t)$ can be defined in terms of matrix elements of
parton light-ray operators $O^{(N)}$ over the hadron states $|P_k\rangle$
with momenta $P_k$: 
\begin{equation}
H^{(N)}(x,\xi,t)=\int\frac{d\lambda}{2\pi}\exp(i\lambda x)\langle
P_{2}|O^{(N)}(\lambda,n)|P_{1}\rangle\,.  \label{F-N-def}
\end{equation}
The light-like vector $n$,
\begin{equation}
n^{2}=0 \,,
\end{equation}
is normalized by the condition
\begin{equation}
n(P_{1}+P_{2})=2\,.
\end{equation}
We use the standard notation of Ji \cite{Ji-98} for parameters $\Delta$, $t$
and $\xi$:
\begin{equation}
\Delta=P_{2}-P_{1}\,,\quad\xi=-\frac{1}{2}(n\Delta),\quad t=\Delta ^{2}\,.
\label{Delta-xi-t-def}
\end{equation}
The definitions of the light-ray operators $O^{(N)}(\lambda,n)$ for various
types of partons are listed in Table \ref{table-N}. 
\begin{table}[tbp]
\begin{tabular}{||l|l|l||}
\hline\hline
Parton & $O^{(N)}(\lambda ,n)$ & $N$ \\ 
\hline\hline
scalar & $\phi ^{\dagger }\left( -\frac{\lambda n
}{2}\right) \phi \left( \frac{\lambda n}{2}\right) $ & $0$ \\ \hline
quark & $\frac12\bar{\psi}\left( -\frac{\lambda n
}{2}\right) (n\cdot \gamma )\psi \left( \frac{\lambda n}{2}\right) $ & $1$
\\ \hline
gluon & $n^{\mu }G_{\mu \nu }^{a}\left( -\frac{
\lambda n}{2}\right) n_{\rho }G^{a,\nu\rho}\left( \frac{\lambda n}{2}\right) 
$ & $2$ \\ \hline\hline
\end{tabular}
\caption{Light-ray operators $O^{(N)}(\protect\lambda ,n)$ for various types
of partons and the corresponding parameter $N$.}
\label{table-N}
\end{table}
We have included the scalar field $\phi$ in this table since the GPD of
the $\phi^{3}$ model will be an essential ingredient of our construction.
The last column of this table contains the number $N$ of factors $n^{\mu}$
appearing in the light-ray operator $O^{(N)}(\lambda,n)$. This number $N$
plays an important role in the formulation of the positivity bounds and
of the
polynomiality conditions and we include $N$ in the notation (\ref{F-N-def})
of the GPD $H^{(N)}(x,\xi,t)$.

In the frame where $(P_{1}+P_{2})^{\perp }=0$ and $n^{\perp }=0$, the
transverse component $\Delta ^{\perp }$ of the hadron momentum transfer
$\Delta $ (\ref{Delta-xi-t-def}) is connected with the parameter $t=\Delta
^{2}$ by the following relation 
\begin{equation}
t=-\frac{|\Delta ^{\perp }|^{2}+4\xi ^{2}M^{2}}{1-\xi ^{2}}\,.
\label{t-Delta-perp}
\end{equation}
Below for the analysis of the positivity bounds we shall need the impact
parameter representation for GPDs
\cite{Burkardt-01,Diehl-02,Burkardt-02-a,Burkardt-02-b,Burkardt-00,Burkardt-02-c}. We define
the GPD in the impact parameter representation as follows:
\[
\tilde{F}^{(N)}\left( x,\xi ,b^{\perp }\right) =\int \frac{d^{2}\Delta
^{\perp }}{(2\pi )^{2}}\exp \left[ i(\Delta ^{\perp }b^{\perp })\right] 
\]
\begin{equation}
\times H^{(N)}\left( x,\xi ,-\frac{|\Delta ^{\perp }|^{2}+4\xi ^{2}M^{2}}{
1-\xi ^{2}}\right) \,.  \label{F-impact-def}
\end{equation}

\section{Polynomiality and positivity}
\label{Poly-posi-section}
\label{polynomiality-positivity-section}

In this section we briefly describe the polynomiality and positivity
properties of GPDs. The polynomiality means that Mellin moments in $x$ of
GPD $H^{(N)}(x,\xi,t)$, 
\begin{equation}
\int_{-1}^{1}dx\,x^{m}H^{(N)}(x,\xi,t)=P_{m+N}(\xi ,t)\,,
\label{polynomiality-power}
\end{equation}
must be polynomials in $\xi$ of degree $m+N$.

The positivity bounds on GPDs have a simple form in the impact parameter
representation (\ref{F-impact-def}).
In Refs.~\cite{Pobylitsa-02-c,Pobylitsa-02-e} the
following inequality was derived:
\[
\,\int_{-1}^{1}d\xi\int_{|\xi|}^{1}dx(1-x)^{-N-4}p^{\ast}
\left( \frac{1-x}{1-\xi}\right) p\left( \frac{1-x}{1+\xi}\right) 
\]
\begin{equation}
\times\tilde{F}^{(N)}\left( x,\xi,\frac{1-x}{1-\xi^{2}}b^{\perp}\right)
\geq0\,.  \label{ineq-q}
\end{equation}
This inequality was derived in Refs.~\cite{Pobylitsa-02-c,Pobylitsa-02-e}
for the case $N=1$
and the generalization for arbitrary $N$ is straightforward.

The inequality (\ref{ineq-q}) should hold for any function $p(z)$. Therefore
we deal with an infinite set of positivity bounds on GPDs. The inequality
(\ref{ineq-q}) (with its generalizations for the nonzero-spin hadrons and for
the full set of the twist-two light-ray operators) covers various
inequalities suggested for GPDs \cite{Martin-98,
Radyushkin-99,PST-99,Ji-98,DFJK-00,Burkardt-01,Pobylitsa-01,Pobylitsa-02,Diehl-02,Burkardt-02-a,Burkardt-02-b}
as particular cases corresponding to some special choice of the function
$p(z)$.

It is well known that the double distribution representation
\cite{MRGDH-94,Radyushkin-96-a,Radyushkin-97} with the $D$ term \cite{PW-99}
\[
H(x,\xi,t)=\int\limits_{|\alpha|+|\beta|\leq1}d\alpha d\beta\delta(x-\xi
\alpha-\beta)\bar{F}_{D}(\alpha,\beta,t) 
\]
\begin{equation}
+\theta(|\xi|-|x|)D\left( \frac{x}{\xi},t\right) \mathrm{sign}(\xi)
\label{DD-representation}
\end{equation}
guarantees the polynomiality property (\ref{polynomiality-power}). Another
interesting parametrization for GPDs supporting the polynomiality was
suggested in Ref.~\cite{PS-02}.

On the other hand, as shown in Ref.~\cite{Pobylitsa-02-d} (see also
Appendix~\ref{r-k-appendix}), the positivity bound on GPDs (\ref{ineq-q}) is
equivalent to the following representation for GPDs in the impact parameter
representation in the region $x>|\xi|$:
\[
\tilde{F}^{(N)}\left( x,\xi,b^{\perp}\right) =(1-x)^{N+1} 
\]
\begin{equation}
\times\sum\limits_{n}Q_{n}\left( \frac{1-x}{1+\xi},(1-\xi)b^{\perp}\right)
Q_{n}\left( \frac{1-x}{1-\xi},(1+\xi)b^{\perp}\right)
\label{pos-representation}
\end{equation}
with arbitrary functions $Q_{n}$. Instead of the discrete summation over $n$
one can use the integration over continuous parameters.

Although both polynomiality and positivity are basic properties that must
hold in any reasonable model of GPDs, usually the model building community
meets a dilemma: one can use the double distribution representation
(\ref{DD-representation}) but it does not guarantee that the infinite set of
inequalities (\ref{ineq-q}) will be satisfied \cite{MMPR-02,TM-02,TM-02-b}.
Alternatively one can build the models based on the representation
(\ref{pos-representation}) or on the so called overlap representation
\cite{DFJK-00}, which also automatically obeys the positivity bounds, but then
one meets problems with the polynomiality. In this paper a rather general
representation for GPDs is suggested which guarantees both positivity and
polynomiality.

\section{General method}
\label{general-method-section}

One could consider the construction of a representation for GPDs which
solves simultaneously positivity and polynomiality constraints as a pure
mathematical problem, looking for functions $Q_{n}$ (\ref{pos-representation})
which allow the double distribution representation (\ref{DD-representation}):
\[
Q_{n}\left( \frac{1-x}{1+\xi },(1-\xi )b^{\perp }\right) Q_{n}\left( 
\frac{1-x}{1-\xi },(1+\xi )b^{\perp }\right) 
\]
\[
=\int \frac{d^{2}\Delta^{\perp }}{(2\pi)^2}\exp \left[ i(\Delta ^{\perp }b^{\perp })\right]
\int\limits_{|\alpha |+|\beta |\leq 1}d\alpha d\beta 
\]
\begin{equation}
\times \delta \left( x-\beta -\xi \alpha \right) F_{D}^{n}\left( \alpha
,\beta ,-\frac{|\Delta ^{\perp }|^{2}+4\xi ^{2}M^{2}}{1-\xi ^{2}}\right) \,.
\label{KK-DD}
\end{equation}
If we manage to find a large set of such functions $Q_{n}$, then taking
linear combinations with positive coefficients we can construct many
solutions of the positivity and polynomiality constraints. This strategy was
used in Ref. \cite{Pobylitsa-02-d}.

On the other hand, the solution of the positivity and polynomiality
constraints is a physical problem and instead of using formal mathematical
methods one can try to solve this problem relying on physical arguments. The
polynomiality and positivity constraints hold in any reasonable quantum
field theory. In particular, we expect these properties in the leading order
perturbative diagrams for GPDs in various field theories. Now it makes sense
to notice that the form of the polynomiality and positivity constraints is
sensitive to the spins of partons and hadrons but not to the dynamics of the
theory. Therefore taking a formal ``superposition'' of the leading-order
perturbative GPDs $H_{\mathcal{M}}(x,\xi,t)$ over various models $\mathcal{M}
$ (and over various values for the parameters of these models) with
arbitrary positive coefficients $c_{\mathcal{M}}$,
\begin{equation}
H(x,\xi,t)=\sum\limits_{\mathcal{M}}c_{\mathcal{M}}H_{\mathcal{M}
}(x,\xi,t)\,,\quad c_{\mathcal{M}}\geq0\,,  \label{model-superposition}
\end{equation}
we also obtain a representation for GPDs which automatically obeys both
polynomiality and positivity constraints.

At first sight the mathematical approach based on relations
(\ref{pos-representation}) and the diagrammatic method (\ref{model-superposition})
are absolutely different ways to solve the positivity and polynomiality
constraints. But there is a deep relation between the two approaches. The
leading order perturbative GPDs $H_{\mathcal{M}}(x,\xi,t)$ obey the positivity
condition (\ref{ineq-q}). Therefore these perturbative GPDs $H_{\mathcal{M}}$
can be represented in the form (\ref{pos-representation}) in the impact
parameter representation (\ref{F-impact-def}). Actually the decomposition
(\ref{pos-representation}) arises automatically if one computes the leading
order triangle Feynman diagrams for $\tilde{F}_{\mathcal{M}
}(x,\xi,b^{\perp}) $ directly in the impact parameter representation \cite
{Cheng-Wu-69}. The sum over $n$ on the right-hand side
(RHS) of Eq. (\ref{pos-representation}) is
nothing else but the sum over the types and polarizations of the
intermediate particles of the triangle graphs for GPDs. Therefore this sum
is \emph{finite} for the leading order perturbative diagrams:
\[
\left. \tilde{F}_{\mathcal{M}}(x,\xi,b^{\perp})\right| _{x>|\xi
|}=(1-x)^{N+1} 
\]
\begin{equation}
\times\sum\limits_{j=1}^{N_{\mathcal{M}}}Q_{\mathcal{M}}^{j}\left( \frac{1-x
}{1+\xi},(1-\xi)b^{\perp}\right) Q_{\mathcal{M}}^{j}\left( \frac{1-x}{1-\xi}
,(1+\xi)b^{\perp}\right) \,.  \label{F-M-Q-general}
\end{equation}
Below we shall explicitly compute this decomposition in the $\phi^{3}$ and
Yukawa models --- see Eqs.~(\ref{F-Y}), (\ref{F-Yukawa}), (\ref{F-Yukawa-5}).

The triangle graph GPDs $H_{\mathcal{M}}(x,\xi,t)$ obey the polynomiality
constraints and automatically have the double distribution representation
(\ref{KK-DD}) which naturally appears in terms of the $\alpha$-parameter
calculation of Feynman diagrams \cite{MRGDH-94,Radyushkin-96-a,Radyushkin-97}.

Now we can combine the physical and mathematical approaches. Triangle graphs
will provide us with functions $Q_{\mathcal{M}}^{j}$ and with the
corresponding double distributions. Taking functions $Q_{\mathcal{M}}^{j}$
generated by the triangle graphs in various models, we can use these
functions $Q_{\mathcal{M}}^{j}$ in the general decomposition
(\ref{pos-representation}). In this way we can construct GPDs obeying both
polynomiality and positivity constraints.

The next step is to take perturbative theories containing several parton
fields with different masses. In this case asymmetric triangle graphs with
different masses will enter the game. Taking the number of different masses
to infinity one arrives at triangle graphs integrated over the masses. Under
certain restrictions on the integration weight this will generate GPDs
satisfying both positivity and polynomiality constraints.

Another important ingredient is the $D$ term (\ref{DD-representation}).
Formally one can use the trick of Ref.~\cite{BMKS-01} and include the $D$
term in the double distribution representation for GPDs. However, for the
analysis of the positivity bounds the explicit form of the $D$ term is much
more convenient. Indeed, the $D$ term vanishes in the region $|x|>|\xi|$ and
therefore it is not restricted by the positivity bound (\ref{ineq-q}). On
the other hand, the $D$ term automatically satisfies the polynomiality
constraint. This means that constructing the solutions of the polynomiality
and positivity constraints we are free to add an arbitrary $D$ term.

\section{Triangle graph in the $\protect\phi^{3}$ model}
\label{triangle-phi-3}

\begin{figure}[ptb]
\begin{center}
\includegraphics[
height=1.8248in,
width=2.9585in
]{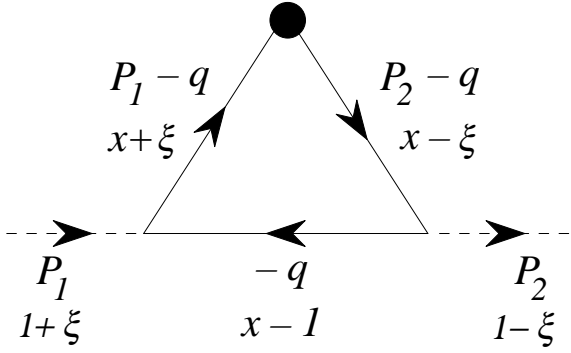}
\end{center}
\caption{The triangle graph for GPDs. The light-cone momentum fractions are
normalized with respect to $(P_1+P_2)/2$.}
\label{fig-1}
\end{figure}

Let us start the analysis of the positivity properties of perturbative
diagrams from the case of scalar partons. The perturbative triangle graph of
the $\phi^{3}$ model of Fig.~\ref{fig-1} is often used as a toy model for
GPDs \cite{Radyushkin-97}. This triangle graph leads to the following
Feynman integral:
\[
H_{\phi^{3}}(x,\xi,t)=(ig_{\phi^{3}})^{2}\int\frac{d^{4}q}{(2\pi)^{4}}
\delta\left[ x-1+\frac{2(nq)}{n(P_{1}+P_{2})}\right] 
\]
\begin{equation}
\times\frac{i}{(P_{1}-q)^{2}-m_{q}^{2}+i0}\frac{i}{q^{2}-m_{q}^{2}+i0}\frac{i
}{(P_{2}-q)^{2}-m_{q}^{2}+i0}\,.  \label{F-phi-3-start}
\end{equation}
Here $g_{\phi^3}$ is the coupling constant of the interaction between
the scalar ``partons'' with the mass $m_{q}$ and the scalar
meson with the mass $M$,
\begin{equation}
P_{1}^{2}=P_{2}^{2}=M^{2}\,.
\end{equation}
It is assumed that the meson stability condition holds: 
\begin{equation}
M\leq2m_{q}\,.  \label{meson-stable}
\end{equation}

We are going to consider the version of the $\phi^{3}$ model with such a
flavor content and couplings which select the diagram of Fig.~\ref{fig-1}
but forbid the cross-channel diagram. The resulting GPD vanishes in the
``antiparton region'' $x<-|\xi|$.

In Appendix~\ref{alpha-calculation} this diagram is computed in the general
case of three different ``parton'' masses in the triangle. Setting these
masses equal in Eq.~(\ref{H-phi3-m123-res})
\begin{equation}
m_1=m_2=m_3=m_q\,,
\end{equation}
we obtain the double distribution representation
for our case (\ref{F-phi-3-start}):
\[
H_{\phi ^{3}}(x,\xi ,t)=\int_{0}^{1}d\alpha
_{1}\int_{0}^{1-\alpha _{1}}d\alpha _{2}F_{\phi ^{3}}^{D}(\alpha
_{1},\alpha _{2},t) 
\]
\begin{equation}
\times \delta \left[ x-1+\alpha _{1}(1+\xi )+\alpha _{2}(1-\xi )\right]
\label{H-phi3-D}
\end{equation}
where
\[
F_{\phi ^{3}}^{D}(\alpha _{1},\alpha _{2},t)=\frac{g_{\phi ^{3}}^{2}}{16\pi
^{2}} 
\]
\begin{equation}
\times \left[ m_{q}^{2}-(\alpha _{1}+\alpha _{2})(1-\alpha _{1}-\alpha
_{2})M^{2}-\alpha _{1}\alpha _{2}t\right] ^{-1}\,.  \label{F-D-phi-3}
\end{equation}
The $\alpha_k$ parameters used here are related to
parameters $\alpha ,\beta $ appearing in Eq.~(\ref{DD-representation})
as follows:
\begin{equation}
\alpha _{1}=\frac{1}{2}(1-\beta -\alpha )\,,\quad \alpha _{2}=\frac{1}{2}
(1-\beta +\alpha )\,.
\end{equation}
In Appendix~\ref{impact-triangle-appendix} the triangle graph is computed in the impact parameter
representation in the region $x>|\xi|$ for the case of three different
``parton'' masses. Setting these masses equal in Eq.~(\ref{tilde-F-res-m123})
we find the impact parameter representation (\ref{F-impact-def})
for our graph at $x>|\xi|$
\[
\tilde{F}_{\phi ^{3}}\left( x,\xi ,b^{\perp }\right) =\frac{1-x}{4\pi } 
\]
\begin{equation}
\times V\left[ \frac{1-x}{1+\xi },(1-\xi )b^{\perp }\right] V\left[ \frac{1-x
}{1-\xi },(1+\xi )b^{\perp }\right]
\label{F-Y}
\end{equation}
with the function $V$ given by  Eq.~(\ref{V-def-123})
in terms of the modified Bessel function $K_0$:
\begin{equation}
V(r,c^{\perp })=\frac{g_{\phi ^{3}}}{2\pi r}K_{0}\left( |c^{\perp }|r^{-1}
\sqrt{m_{q}^{2}-r\left( 1-r\right) M^{2}}\right) \,.  \label{Y-def}
\end{equation}

The factorized form of the result (\ref{F-Y}) for the GPD in the impact parameter representation
obtained in the $\phi ^{3}$ model is an illustration of the general
decomposition of triangle diagrams (\ref{F-M-Q-general}). We see that
in our case the sum on the RHS of Eq.~(\ref{F-M-Q-general})
contains only one term. The reason is that the $q$ propagator
of our diagram corresponds to a spin-zero particle.

Introducing the variables (see Appendix~\ref{r-k-appendix} for more
details)
\begin{equation}
r_{1}=\frac{1-x}{1+\xi },\quad r_{2}=\frac{1-x}{1-\xi }\,
\label{r1-r2-def-0}
\end{equation}
instead of $x$, $\xi $, and working in the region $x>|\xi |$ (i.e.
$0<r_{1},r_{2}<1$), one can rewrite Eq.~(\ref{F-Y})
in the form
\[
\tilde{F}_{\phi^{3}}\left( x,\xi,\frac{1-x}{1-\xi^{2}}b^{\perp}\right)
\]
\begin{equation}
 =\frac{1}{2\pi}\frac{r_{1}r_{2}}{r_{1}+r_{2}}V(r_{1},r_{1}b^{
\perp})V(r_{2},r_{2}b^{\perp})
\qquad(x>|\xi|)
\,.
\end{equation}

\begin{widetext}
\section{Triangle graph in Yukawa model}
\label{triangle-Yukawa-section}
Now let us compute the ``quark-in-meson'' GPD in Yukawa model. The same
triangle graph of Fig.~\ref{fig-1} (now with the fermion loop) leads to the
following Feynman integral
\[
H_{Y}(x,\xi,t)=\frac{1}{2}g_{Y}^{2}\int\frac{d^{4}q}{(2\pi)^{4}}
\delta\left[ x-1+\frac{2(nq)}{n(P_{1}+P_{2})}\right] 
\]
\begin{equation}
\times
\mathrm{Tr}\left[ (n\gamma)\frac{i}{(P_{1}-q)\gamma-m_{q}+i0}\frac{i}{
-(q\gamma)-m_{q}+i0}\frac{i}{(P_{2}-q)\gamma-m_{q}+i0}\right]\,.
\end{equation}
\begin{figure}[ptb]
\begin{center}
\includegraphics[
height=2.1577in,
width=4.6709in
]{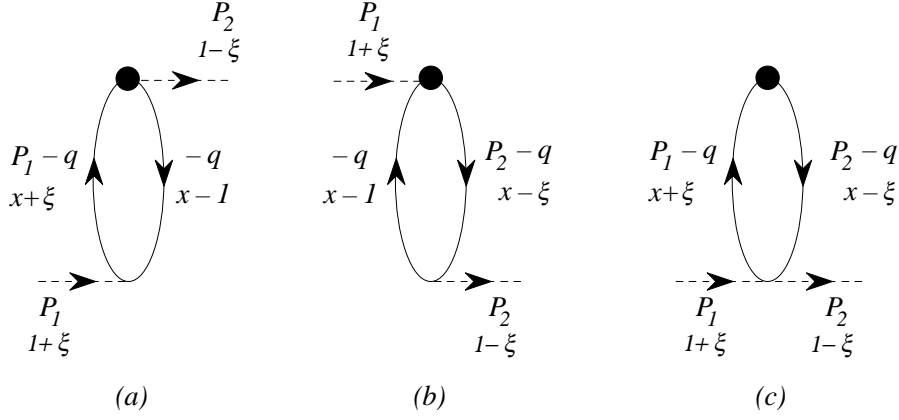}
\end{center}
\caption{Reduced diagrams coming from the triangle graph in Yukawa model.}
\label{fig-2}
\end{figure}%
The factor of $1/2$ on the RHS is inherited from the light-ray fermion
operator $O^{(1)}$ (see Table~\ref{table-N}), and $g_Y$ is the coupling constant.
Again we assume that the
flavor content and couplings are chosen so that the cross-channel triangle
diagram is forbidden so that we deal with the GPD vanishing at $x<-|\xi|$.

The trace of the Dirac matrices can be represented in the following form:
\[
\mathrm{Tr}\left\{ (n\gamma)\left[ (P_{1}-q)\gamma+m_{q}\right] \left[
-(q\gamma)+m_{q}\right] \left[ (P_{2}-q)\gamma+m_{q}\right] \right\} 
=4\left\{ \left[ n\left( -q+\frac{1}{2}P_{1}+\frac{1}{2}P_{2}\right) \right]
(q^{2}-m_{q}^{2})\right. 
\]
\begin{equation}
+\frac{1}{2}(nP_{1})\left[ (P_{2}-q)^{2}-m_{q}^{2}\right] +\frac{1}{2}
(nP_{2})\left[ (P_{1}-q)^{2}-m_{q}^{2}\right] 
\left. -(nq)\left[ 4m_{q}^{2}-(P_{1}P_{2})\right] -\left[ n\left(
P_{1}+P_{2}\right) \right] \left( \frac{1}{2}M^{2}-2m_{q}^{2}\right)
\right\} \,.  \label{Yukawa-reduction}
\end{equation}

Most of the terms on the RHS of Eq.~(\ref{Yukawa-reduction}) contain factors
which cancel one of the propagators in the denominator so that one arrives
at reduced diagrams containing only two propagators (Fig.~\ref{fig-2}). The
contribution of the nonreduced part is
\[
H_{Y}(x,\xi,t)=2g_{Y}^{2}\int\frac{d^{4}q}{(2\pi)^{4}}
\delta\left[ x-1+\frac{2(nq)}{n(P_{1}+P_{2})}\right] 
\left\{\ -(nq)\left[ 4m_{q}^{2}-(P_{1}P_{2})\right] -\left[ n\left(
P_{1}+P_{2}\right) \right] \left( \frac{1}{2}M^{2}-2m_{q}^{2}\right) \right\}
\]
\begin{equation}
\times\frac{i}{(P_{1}-q)^{2}-m_{q}^{2}+i0}\frac{i}{q^{2}-m_{q}^{2}+i0}\frac{i
}{(P_{2}-q)^{2}-m_{q}^{2}+i0} 
+\mathrm{reduced\,diagrams}\,.
\end{equation}
Comparing the RHS with Eq.~(\ref{F-phi-3-start}), we see that we have reduced
the calculation of the GPD in the Yukawa model to the scalar GPD in the $\phi^{3}$
model
\begin{equation}
H_{Y}(x,\xi,t)=2\frac{g_{Y}^{2}}{g_{\phi^{3}}^{2}}\left[ \frac{1}{2}
(1-x)t-x\left( 4m_{q}^{2}-M^{2}\right) \right] H_{\phi^{3}}(x,\xi,t) 
+\mathrm{reduced\,diagrams}\,.  \label{H-Y-H-phi}
\end{equation}
The reduced diagrams \emph{(a)} and \emph{(b)} of Fig.~\ref{fig-2} give a $t$
independent contribution. Therefore in the impact parameter representation
they vanish if $b^{\perp}\neq0$. The contribution of the reduced diagram 
\emph{(c)} of Fig.~\ref{fig-2} has the structure of a $D$ term
(\ref{DD-representation}) which vanishes if $|x|>|\xi|$. Thus all three
reduced diagrams can be ignored if one is interested in the region
$|x|>|\xi|$, $b^{\perp}\neq0$.

Let us transform Eq.~(\ref{H-Y-H-phi}) into the impact parameter
representation omitting the reduced diagrams. Parameter $t$
(\ref{t-Delta-perp}) becomes a differential operator:
\begin{equation}
t=-\frac{|\Delta^{\perp}|^{2}+4\xi^{2}M^{2}}{1-\xi^{2}}\rightarrow \frac{1}{
1-\xi^{2}}\left[ \left( \frac{\partial}{\partial b^{\perp}}\right)
^{2}-4\xi^{2}M^{2}\right] \,.
\end{equation}
With this expression for $t$ and with the representation (\ref{F-Y}) for
$\tilde{F}_{\phi^{3}}(x,\xi,b^{\perp})$ we find from Eq.~(\ref{H-Y-H-phi})
\[
\left. \tilde{F}_{Y}(x,\xi,b^{\perp})\right| _{b^{\perp}\neq0,x>|\xi |}=
\frac{g_{Y}^{2}}{g_{\phi^{3}}^{2}}\frac{1-x}{2\pi} 
\left\{ (1-x)\frac{1}{2}\frac{1}{1-\xi^{2}}\left[ \left( \frac{\partial
}{\partial b^{\perp}}\right) ^{2}-4\xi^{2}M^{2}\right] -x\left(
4m_{q}^{2}-M^{2}\right) \right\} 
\]
\begin{equation}
\times V\left[ \frac{1-x}{1+\xi},(1-\xi)b^{\perp}\right] V\left[ \frac{1-x}{
1-\xi},(1+\xi)b^{\perp}\right] \,.  \label{F-V-Yukawa}
\end{equation}
Functions $V\left( r,c^{\perp}\right) $ expressed in terms of the modified
Bessel functions (\ref{Y-def}) obey the following differential equation
\begin{equation}
\left( \frac{\partial}{\partial c^{\perp}}\right) ^{2}V\left( r,c^{\perp
}\right) =\left[ \left( rM^{2}+r^{-1}m_{q}^{2}\right) -M^{2}\right]
r^{-1}V\left( r,c^{\perp}\right) \,.
\end{equation}
Using this differential equation and variables $r_{k}$ (\ref{r1-r2-def-0})
we find from Eq.~(\ref{F-V-Yukawa})
\[
\left. \tilde{F}_{Y}\left( x,\xi,\frac{1-x}{1-\xi^{2}}b^{\perp}\right)
\right| _{b^{\perp}\neq0,\,x>|\xi|} 
=(1-x)^{2}\frac{g_{Y}^{2}}{g_{\phi^{3}}^{2}}\frac{1}{2\pi r_{1}r_{2}}\left\{
m_{q}^{2}\left[ (1-2r_{1})V(r_{1},r_{1}b^{\perp})\right] \left[
(1-2r_{2})V(r_{2},r_{2}b^{\perp})\right] \right. 
\]
\begin{equation}
\left. +\left[ \nabla_{b^{\perp}}V\left( r_{1},r_{1}b^{\perp}\right) \right] 
\left[ \nabla_{b^{\perp}}V\left( r_{2},r_{2}b^{\perp}\right) \right]
\right\} \,.  \label{F-Yukawa}
\end{equation}
We see that the RHS has the general structure (\ref{F-M-Q-general})
satisfying the positivity bounds.

One can also compute the GPD $H_{Y,\gamma_{5}}(x,\xi,t)$ in the Yukawa model
with the pseudoscalar coupling. Replacing the interaction
$g_{Y}\phi\bar{\psi}\psi\rightarrow g_{Y}\phi\bar{\psi}i\gamma_{5}\psi$,
one slightly changes the Dirac trace (\ref{Yukawa-reduction}), which leads to
the following modification of the GPD:
\begin{equation}
H_{Y,\gamma_{5}}(x,\xi,t)=H_{Y}(x,\xi,t)+8m_{q}^{2}\frac{g_{Y}^{2}}{
g_{\phi^{3}}^{2}}xH_{\phi^{3}}(x,\xi,t)\,.  \label{scal-pseudo-Yukawa-F}
\end{equation}
In the impact parameter representation we again find an example of the
general structure (\ref{F-M-Q-general}) which guarantees the positivity: 
\[
\left. \tilde{F}_{Y,\gamma_{5}}\left( x,\xi,\frac{1-x}{1-\xi^{2}}b^{\perp
}\right) \right| _{b^{\perp}\neq0,x>|\xi|}=\tilde{F}_{Y}\left( x,\xi,\frac{
1-x}{1-\xi^{2}}b^{\perp}\right) 
+8m_{q}^{2}x\frac{g_{Y}^{2}}{g_{\phi^{3}}^{2}}\tilde{F}_{\phi^{3}}\left(
x,\xi,\frac{1-x}{1-\xi^{2}}b^{\perp}\right) 
\]
\begin{equation}
=(1-x)^{2}\frac{g_{Y}^{2}}{g_{\phi^{3}}^{2}}\frac{1}{2\pi r_{1}r_{2}}\left\{
m_{q}^{2}V(r_{1},r_{1}b^{\perp})V(r_{2},r_{2}b^{\perp})\right. 
+\left. \left[ \nabla_{b^{\perp}}V\left( r_{1},r_{1}b^{\perp}\right) \right] 
\left[ \nabla_{b^{\perp}}V\left( r_{2},r_{2}b^{\perp}\right) \right]
\right\} \,.  \label{F-Yukawa-5}
\end{equation}

\end{widetext}

\section{Building models of GPDs from triangle graphs}
\label{building-section}

In the previous section we have explicitly checked that the triangle
diagrams in Yukawa model generate GPDs satisfying both polynomiality and
positivity constraints. Since the fermion-in-(pseudo)scalar GPDs obey the
same polynomiality and positivity constraints in the Yukawa model and in QCD we
can use the triangle GPDs of the Yukawa model as elements for the construction
of models of the quark GPD in pion, compatible with the polynomiality and
positivity constraints.

The first step is to mix the GPDs of the scalar and pseudoscalar Yukawa model
\begin{equation}
C_{1}H_{Y}(x,\xi,t)+C_{2}H_{Y,\gamma_{5}}(x,\xi,t)  \label{Yukawa-mixed}
\end{equation}
with positive coefficients, which is equivalent to the Yukawa model with the
coupling
\begin{equation}
g_{Y}\phi\bar{\psi}\left( \sqrt{C_{1}}+i\gamma_{5}\sqrt{C_{2}}\right) \psi\,.
\end{equation}
Now let us show that the function
\begin{equation}
(1-x)H_{\phi^{3}}(x,\xi,t)  \label{one-x-trick}
\end{equation}
also obeys both polynomiality and positivity constraints for the quark GPD
in pion. Indeed, the positivity inequality (\ref{ineq-q}) for the
fermion-in-scalar GPD ($N=1$) differs from the case of the scalar-in-scalar
GPD ($N=0$) exactly by the factor of $(1-x)$. The polynomiality condition
(\ref{polynomiality-power}) for the fermion-in-scalar GPD also allows one
more degree of $x$ compared to the GPD in the $\phi^{3}$ model. Now we can
use all available elements, (\ref{Yukawa-mixed}) and (\ref{one-x-trick}), to
build models for the pion GPD. For any positive coefficients $C_{k}$ the
following combination will satisfy both polynomiality and positivity
constraints: 
\begin{widetext}
\[
C_{1}H_{Y}(x,\xi,t)+C_{2}H_{Y,\gamma_{5}}(x,\xi,t)+C_{3}(1-x)
H_{\phi^{3}}(x,\xi,t) 
\]
\begin{equation}
=2\frac{g_{Y}^{2}}{g_{\phi^{3}}^{2}}\left\{ (C_{1}+C_{2})\left[ \frac{1}{2}
(1-x)t-x\left( 4m_{q}^{2}-M^{2}\right) \right] \right. 
\left. +4C_{2}m_{q}^{2}x+C_{3}\frac{g_{\phi^{3}}^{2}}{2g_{Y}^{2}}
(1-x)\right\} H_{\phi^{3}}(x,\xi,t)\,.  \label{var-H-phi3}
\end{equation}

The next step is to consider triangle graphs with arbitrary masses. We start
from the $\phi^{3}$ model. Let us take the triangle graph of Fig.~\ref{fig-1}
with the masses $m_{1}$ for the $(P_{1}-q)$-propagator, $m_{2}$ for
$(P_{2}-q)$ and $m_{3}$ for $q$. This graph is computed in
Appendix~\ref{alpha-calculation}. The result (\ref{H-phi3-m123-res})
can be represented in the following form:
\begin{equation}
H_{\phi^{3}}(x,\xi,t|m_{1},m_{2},m_{3})=\int_{0}^{1}d\alpha_{1}\int
_{0}^{1-\alpha_{1}}d\alpha_{2}F_{\phi^{3}}^{D}(\alpha_{1},
\alpha_{2},t|m_{1},m_{2},m_{3}) 
\delta\left[ x-1+\alpha_{1}(1+\xi)+\alpha_{2}(1-\xi)\right] \,,
\label{H-FD-phi3}
\end{equation}
\[
F_{\phi^{3}}^{D}(\alpha_{1},\alpha_{2},t|m_{1},m_{2},m_{3}) 
=\frac{g_{\phi^{3}}^{2}}{16\pi^{2}}\left\{ -\alpha_{1}\alpha_{2}t-(\alpha
_{1}+\alpha_{2})\left[ 1-(\alpha_{1}+\alpha_{2})\right] M^{2}\right. 
\]
\begin{equation}
\left. +\left[ \alpha_{1}m_{1}^{2}+\alpha_{2}m_{2}^{2}+(1-\alpha_{1}-
\alpha_{2})m_{3}^{2}\right] \right\} ^{-1}\,.  \label{q-phi3-res-alpha}
\end{equation}
The corresponding impact parameter representation in the region $x>|\xi|$
computed in Appendix~\ref{impact-triangle-appendix} is given by
Eq.~(\ref{tilde-F-res-m123}):
\begin{equation}
\tilde{F}_{\phi^{3}}\left( \left. x,\xi,\frac{1-x}{1-\xi^{2}}b^{\perp
}\right| m_{1},m_{2},m_{3}\right) 
=\frac{1}{4\pi}(1-x)V(r_{1},r_{1}b^{\perp}|m_{1},m_{3})V(r_{2},r_{2}b^{\perp
}|m_{2},m_{3})\,.  \label{F-Y-3m}
\end{equation}
The function $V$ is defined by Eq.~(\ref{V-def-123}):
\begin{equation}
V(r_{1},c^{\perp}|m_{1},m_{3}) 
=\frac{g_{\phi^{3}}}{2\pi r_{1}}K_{0}\left( |c^{\perp}|r_{1}^{-1}\sqrt {
r_{1}m_{1}^{2} +(1-r_{1})m_{3}^{2} -r_{1}\left( 1- r_{1}\right) M^{2}}
\right) \,.
\end{equation}
\end{widetext}

Any single triangle graph automatically satisfies the polynomiality
constraint. Therefore mixing the contributions of various triangle graphs we
must take care only about the positivity. Keeping in mind the factorized
structure (\ref{F-Y-3m}) we see that the integral
\[
H^{(0)}(x,\xi,t)=\int dm_{1}\int dm_{2}\int dm_{3} 
\]
\begin{equation}
\times s(m_{1},m_{2},m_{3})H_{\phi^{3}}(x,\xi,t|m_{1},m_{2},m_{3})
\label{H-0-phi}
\end{equation}
is compatible with the positivity if the weight $s$ has the structure
\begin{equation}
s(m_{1},m_{2},m_{3})=\int d\lambda
a(m_{1},m_{3},\lambda)a^{\ast}(m_{2},m_{3},\lambda)  \label{sigma-s}
\end{equation}
where the function $a(m_{1},m_{3},\lambda)$ is arbitrary. This
representation is equivalent to the following property
of $s(m_{1},m_{2},m_{3})$:
\begin{equation}
\int dm_{1}\int dm_{2}s(m_{1},m_{2},m_{3})f(m_{1})f^{\ast}(m_{2})\geq0
\end{equation}
for any function $f(m)$ and for any value of $m_{3}$.
Since we are interested in real GPDs even in $\xi$, we must work
with real functions $a_k$.

It is also assumed that functions $a(m_{1},m_{3},\lambda)$ are compatible
with the stability of the meson:
\begin{equation}
a(m_{1},m_{3},\lambda)=0\quad {\rm if}\quad m_1+m_3<M\,.
\label{meson-stable-a}
\end{equation}

Now we can turn to Yukawa model, generalize the representation
(\ref{var-H-phi3}) to the case of different masses $m_{1},m_{2},m_{3}$ and
integrate over these masses by analogy with the $\phi^{3}$ model,
Eq.~(\ref{H-0-phi}). The generalization of Eqs.~(\ref{H-Y-H-phi}) and
(\ref{scal-pseudo-Yukawa-F}) for the case of different masses $m_k$ is
\begin{widetext}
\[
H_{Y}(x,\xi ,t|m_{1},m_{2},m_{3})=\left( \frac{g_{Y}}{g_{\phi ^{3}}}\right)^{2}
\left\{
(1-x)t +2xM^{2} -2x(m_{1}+m_{3})(m_{2}+m_{3}) -(m_{1}-m_{2})^{2} \right. 
\]
\begin{equation}
\left. +\xi (m_{1}-m_{2})(m_{1}+m_{2}+2m_{3})\right\} H_{\phi
^{3}}(x,\xi ,t|m_{1},m_{2},m_{3})+\mathrm{reduced\,diagrams}\,,
\label{HY-m123}
\end{equation}
\begin{equation}
H_{Y,\gamma _{5}}(x,\xi ,t|m_{1},m_{2},m_{3})=H_{Y}(x,\xi ,t)+4\left( \frac{
g_{Y}}{g_{\phi ^{3}}}\right) ^{2}m_{3}\left[ (x-\xi )m_{1}+(x+\xi )m_{2}
\right] H_{\phi ^{3}}(x,\xi ,t|m_{1},m_{2},m_{3})\,.  \label{HY5-m123}
\end{equation}
Combining the ``superposition''
\[
\int dm_{1}\int dm_{2}\int dm_{3}\Biggl\{
\left( \frac{g_{\phi ^{3}}}{g_{Y}}\right)
^{2}\left[ s_{1}(m_{1},m_{2},m_{3})H_{Y}(x,\xi
,t|m_{1},m_{2},m_{3})+s_{2}(m_{1},m_{2},m_{3})H_{Y,\gamma ^{5}}(x,\xi
,t|m_{1},m_{2},m_{3})\right] 
\]
\begin{equation}
+(1-x)s_{2}(m_{1},m_{2},m_{3})H_{\phi ^{3}}(x,\xi ,t|m_{1},m_{2},m_{3})
\Biggr\}
\end{equation}
with a $D$ term and using Eqs.~(\ref{HY-m123}), (\ref{HY5-m123}),
we arrive at the following solution of the positivity and
polynomiality constraints for the
fermion-in-scalar GPDs ($H^{(N)}$ with $N=1$):
\[
H^{(1)}\left( x,\xi,t\right) =\int dm_{1}\int dm_{2}\int dm_{3} 
H_{\phi^{3}}\left(x,\xi,t|m_{1},m_{2},m_{3}\right) 
\left\{ \left[ s_{1}(m_{1},m_{2},m_{3})+s_{2}(m_{1},m_{2},m_{3})\right]
\right. 
\]
\[
\times\left[ (1-x)t+2x M^{2}-2x(m_{1}+m_{3})(m_{2}+m_{3}) \right. 
\left. -(m_{1}-m_{2})^{2}+\xi(m_{1}-m_{2})(m_{1}+2m_{2}+2m_{3})\right] 
\]
\begin{equation}
+4s_{2}(m_{1},m_{2},m_{3})m_{3}\left[ (x-\xi)m_{1}+(x+\xi)m_{2}\right] 
\left. +s_{3}(m_{1},m_{2},m_{3})(1-x)\right\}
+\theta\left( |\xi|-|x|\right) D\left( \frac{x}{\xi},t\right) \mathrm{sign}
(\xi)\,.  \label{H-mixture}
\end{equation}
\end{widetext}
Here the integration weights $s_{k}$ must have the structure
\begin{equation}
s_{k}(m_{1},m_{2},m_{3})=\int d\lambda a_{k}(m_{1},m_{3},\lambda)a_{k}^{\ast
}(m_{2},m_{3},\lambda)  \label{s-from-c}
\end{equation}
with arbitrary functions $a_{k}(m_{1},m_{3},\lambda)$ obeying the
meson stability condition (\ref{meson-stable-a}). The functions $a_k$ must be real
if one is interested in real $\xi$-even GPDs. We remind the reader that the
term $D(x/\xi,t)$ on the RHS of Eq.~(\ref{H-mixture}) is not constrained by
the polynomiality and positivity.

The triangle GPD $H_{\phi^{3}}$ vanishes in the antiquark region $x<-|\xi|$.
Therefore the construction (\ref{H-mixture}) should be modified by adding
a similar contribution with the replacement $x\rightarrow-x$
and with its own set of coefficients $s_{k}$.

As mentioned above, we have checked that the GPDs obtained from the triangle
graphs satisfy the positivity bounds in the impact parameter representation
only at $b^{\perp}\neq0$. At $b^{\perp}=0$ we must take into account
the $\delta(b^\perp)$ contributions coming
from the reduced diagrams \emph{(a), (b)} of Fig.~\ref{fig-2}, which depend
on the normalization point $\mu$ and can violate the positivity bounds. This
$b^{\perp }=0$ singularity of the triangle diagrams is the leading order
perturbative manifestation of more serious problems which can be met due to
a nontrivial interplay between the two scales $\mu^{-1}$ and $b^{\perp}$ 
\cite{Diehl-02}. If one wants to construct models of GPDs avoiding this
small $b^{\perp}$ problem, then one can impose the following condition on the
coefficients $a_{k}(m_{1},m_{3},\lambda)$ appearing in our construction of
the integration weight (\ref{s-from-c})
\begin{equation}
\int dm_{1}a_{k}(m_{1},m_{3},\lambda)=0 \qquad(k=1,2) \,.
\label{c-constraint}
\end{equation}
Indeed, the reduced diagram of Fig.~\ref{fig-2} \emph{(b)} does not depend
on $m_{1}$, therefore after the integration over the masses in
Eq.~(\ref{H-mixture}) with the weight (\ref{s-from-c}) obeying the condition
(\ref{c-constraint}), the contribution of the diagram \emph{(b)} vanishes. The
contribution of the diagram \emph{(a)} is $m_{2}$ independent and vanishes
after the integration over $m_{2}$. Condition (\ref{c-constraint}) also
suppresses the unacceptable large $t$ behavior of triangle graphs. Note that
Eq.~(\ref{c-constraint}) means that the functions $a_{k}(m_{1},m_{3},\lambda)$
cannot be positive everywhere. This is not a problem because in order to
satisfy the positivity bounds on GPDs we need only the construction
(\ref{s-from-c}) for the functions $s_{k}$ and we have no restrictions on the sign of
the functions $a_{k}$.

\section{Positivity of forward parton distributions}
\label{FPD-section}

The positivity of forward parton distributions (FPDs) is a consequence of
the positivity bounds on GPDs. This idea is present in an explicit or
implicit form practically in all papers dealing with the positivity bounds
on GPDs \cite{Martin-98,Radyushkin-99,PST-99,Ji-98,
DFJK-00,Burkardt-01,Pobylitsa-01,Pobylitsa-02,Diehl-02,Burkardt-02-a,Burkardt-02-b,Pobylitsa-02-c,Pobylitsa-02-e}
. Since our construction of scalar-in-scalar GPDs (\ref{H-0-phi}) and
fermion-in-scalar GPDs (\ref{H-mixture}) satisfies the positivity bounds on
GPDs (\ref{ineq-q}), the positivity of the corresponding FPDs is
predetermined. Nevertheless the direct explicit check of the positivity
of FPDs is
rather interesting. In particular, in the case the fermion-in-scalar GPDs
made of the triangle graphs of the Yukawa model, the analysis of the forward
limit is instructive for understanding the role of the
``divergence-cancellation'' condition (\ref{c-constraint}).

\begin{widetext}
The FPD of the $\phi^{3}$ model is given by
\begin{equation}
q_{\phi^{3}}(x|m_{1},m_{2},m_{3})=\left.
H_{\phi^{3}}(x,\xi,t|m_{1},m_{2},m_{3})\right| _{\xi=0,\,t=0} 
=\frac{g_{\phi^{3}}^{2}}{16\pi^{2}}\frac{\theta(x)}{m_{1}^{2}-m_{2}^{2}}\ln
\frac{xm_{3}^{2}+(1-x)m_{1}^{2}-x(1-x)M^{2}}{
xm_{3}^{2}+(1-x)m_{2}^{2}-x(1-x)M^{2}}\,.  
\label{q-FPD-phi-3-again}
\end{equation}
This result can be obtained by taking the forward limit in
Eq.~(\ref{H-FD-phi3}). Note that
here we define the FPD $q_{\phi^{3}}$ as the forward limit of the GPD
$H_{\phi^{3}}$. This differs by a factor of $x^{-1}$ from the physical
definition in terms of the density of partons. If we take the FPD $q_{\phi
^{3}}$ for the $\phi^{3}$ model without integrating this FPD over masses
$m_{k}$, then in the symmetric case $m_{1}=m_{2}$ we find from
Eq.~(\ref{q-FPD-phi-3-again})
\begin{equation}
q_{\phi^{3}}(x|m_{1},m_{1},m_{3})
=\frac{g_{\phi^{3}}^{2}} {
16\pi^{2}} 
\frac{(1-x)\theta(x)}{xm_{3}^{2}+(1-x)m_{1}^{2}-x(1-x)M^{2}}\,.
\end{equation}
The positivity of this function is obvious if we assume the meson stability
condition
\begin{equation}
M\leq m_{1}+m_{3}\,.  \label{meson-stability-1-3}
\end{equation}
Indeed,
\begin{equation}
xm_{3}^{2}+(1-x)m_{1}^{2}-x(1-x)M^{2} 
=\left[ xm_{3}-(1-x)m_{1}\right] ^{2}+x(1-x)\left[ (m_{1}+m_{3})^{2}-M^{2}
\right] \geq0\,.
\end{equation}
Next, we can consider the scalar-in-scalar GPDs constructed according to Eq.
(\ref{H-0-phi}) and take the forward limit. With expressions (\ref{sigma-s})
for $s$ and (\ref{q-FPD-phi-3-again}) for $q_{\phi^{3}}$ we find
\[
q(x)=\int dm_{1}\int dm_{2}\int
dm_{3}s(m_{1},m_{2},m_{3})q_{\phi^{3}}(x|m_{1},m_{2},m_{3})
\]
\begin{equation}
=\int d\lambda\int dm_{1}\int dm_{2}\int dm_{3}
 a(m_{1},m_{3},\lambda)a^{\ast
}(m_{2},m_{3},\lambda) 
\frac{g_{\phi^{3}}^{2}}{16\pi^{2}}\frac{\theta(x)}{m_{1}^{2}-m_{2}^{2}}
\ln\frac{xm_{3}^{2}+(1-x)m_{1}^{2}-x(1-x)M^{2}}{
xm_{3}^{2}+(1-x)m_{2}^{2}-x(1-x)M^{2}}\,.
\end{equation}
The positivity of this expression is a consequence of the following
inequality valid for any function $f$ and for any $\nu$ (see
Appendix~\ref{inequalities-appendix}) 
\begin{equation}
\int\limits_{m_{1}^{2}+\nu>0}dm_{1}\int\limits_{m_{2}^{2}+
\nu>0}dm_{2}f(m_{1})f^{\ast}(m_{2})\frac{1}{m_{1}^{2}-m_{2}^{2}}\ln\frac{
m_{1}^{2}+\nu}{m_{2}^{2}+\nu}\geq0\,.  \label{log-ineq-1}
\end{equation}

Now let us turn to the Yukawa model. A straightforward calculation allows us to
express the triangle graph contribution (with different parton masses) to
the FPD of Yukawa model $q_{Y}$ in terms of the triangle FPD $q_{\phi^{3}}$
of the $\phi^{3}$ model as follows:
\[
q_{Y}(x|m_{1},m_{2},m_{3})=\left( \frac{g_{Y}}{g_{\phi^{3}}}\right) ^{2} 
\left[ 
2xM^{2}
-2x(m_{1}+m_{3})(m_{2}+m_{3})
-(m_{1}-m_{2})^{2}
\right]
q_{\phi^{3}}(x) 
\]
\begin{equation}
-\frac{g_{Y}^{2}\theta(x)}{16\pi^{2}}\left\{ \ln\frac{
(1-x)m_{1}^{2}+xm_{3}^{2}-x(1-x)M^{2}}{\mu^{2}}\right. 
+\left. \ln\frac{(1-x)m_{2}^{2}+xm_{3}^{2}-x(1-x)M^{2}}{\mu^{2}}\right\} \,.
\label{q-Yukawa}
\end{equation}
The first term on the RHS containing $q_{\phi^3}$ can be obtained by taking the forward
limit in Eq.~(\ref{HY-m123}). The logarithmic terms on the RHS
are generated by the reduced diagrams {\em (a), (b)} of Fig.~\ref{fig-2}.
The ultraviolet divergences of the Yukawa model are renormalized at the scale
$\mu$. For fixed parton masses, the FPD $q_{Y}$ (\ref{q-Yukawa}) depends on
the normalization point $\mu$ via the additive term $\ln\mu$. This simple
$\mu$ dependence obviously leads to the violation of the positivity at low
normalization points and to the restoration of the positivity at large $\mu$
(here the formal behavior of the triangle graph is meant and not the
properties of the full Yukawa model).

Next we want to study the forward limit of the fermion-in-scalar GPDs
constructed according to Eq.~(\ref{H-mixture}). Since the logarithmic $\mu$
dependent terms in Eq.~(\ref{q-Yukawa}) depend either on $m_{1}$ or on
$m_{2} $ but not on both $m_{1}$ and $m_{2}$ simultaneously, we conclude that
these logarithmic terms will be cancelled by the integration
over $m_{1}$ and $m_{2}$ due to the condition (\ref{c-constraint}). For
simplicity let us consider the case $a_{2}=a_{3}=0$. Then Eq.~(\ref{H-mixture})
generates the following FPD:
\[
\int dm_{1}\int dm_{2}\int
dm_{3}s_{1}(m_{1},m_{2},m_{3})q_{Y}(x|m_{1},m_{2},m_{3}) 
=\frac{g_{Y}^{2}}{16\pi^{2}}\theta(x)\int d\lambda\int dm_{1}\int dm_{2}\int
dm_{3}a_{1}(m_{1},m_{3},\lambda)
\]
\begin{equation}
\times
a_{1}^{\ast}(m_{2},m_{3},\lambda) 
\left[
2xM^{2}
-2x(m_{1}+m_{3})(m_{2}+m_{3})
-(m_{1}-m_{2})^{2}
\right] 
\frac{1}{m_{1}^{2}-m_{2}^{2}}\ln\frac{
xm_{3}^{2}+(1-x)m_{1}^{2}-x(1-x)M^{2}}{xm_{3}^{2}+(1-x)m_{2}^{2}-x(1-x)M^{2}}
\,.
\end{equation}
The positivity of this FPD reduces to the following inequality:
\[
\int d\lambda\int dm_{1}\int dm_{2}\int dm_{3}a_{1}(m_{1},m_{3},\lambda
)a_{1}^{\ast}(m_{2},m_{3},\lambda)  \frac{1}{m_{1}^{2}-m_{2}^{2}}
\]
\begin{equation}
\times\left[
2xM^{2}
-2x(m_{1}+m_{3})(m_{2}+m_{3})
 -(m_{1}-m_{2})^{2}
\right] 
\ln\frac{xm_{3}^{2}+(1-x)m_{1}^{2}-x(1-x)M^{2}}{
xm_{3}^{2}+(1-x)m_{2}^{2}-x(1-x)M^{2}}\geq0\,.
\label{FPD-Y-ineq}
\end{equation}
\end{widetext}
Fixing $\lambda$ and $m_{3}$, one can show that this inequality holds already
after the integration over $m_{1}$ and $m_{2}$. In order to see this, we
first have to rearrange the factor in the brackets as follows:
\[
2xM^{2} 
-2x(m_{1}+m_{3})(m_{2}+m_{3})
-(m_{1}-m_{2})^{2}
\]
\begin{equation}
=A_{1}+A_{2}
\end{equation}
where
\begin{equation}
A_{1}=2(1-x)\left( m_{1}-\frac{x}{1-x}m_{3}\right) \left( m_{2}-\frac{x}{1-x}
m_{3}\right) \,,
\end{equation}
\begin{equation}
A_{2}=-\left[ m_{1}^{2}+m_{2}^{2}+2x\left( \frac{m_{3}^{2}}{1-x}
-M^{2}\right) \right] \,.
\end{equation}
The positivity of the $A_{1}$ contribution to the inequality~(\ref{FPD-Y-ineq})
follows from
the inequality (\ref{log-ineq-1}) combined with the meson stability condition
(\ref{meson-stable-a}) for the function $a_1$.
In order to prove the positivity of the
contribution of $A_{2}$ to the inequality (\ref{FPD-Y-ineq}),
one has to use the following inequality derived in
Appendix~\ref{inequalities-appendix}: 
\[
\int\limits_{m_{1}^{2}+\nu>0}dm_{1}f(m_{1})\int\limits_{m_{2}^{2}+
\nu>0}dm_{2}f^{\ast}(m_{2}) 
\]
\begin{equation}
\times\frac{m_{1}^{2}+m_{2}^{2}+2\nu}{m_{1}^{2}-m_{2}^{2}}\ln\frac{
m_{1}^{2}+\nu}{m_{2}^{2}+\nu}\leq0\,,
\end{equation}
This inequality holds for any function $f$ obeying the condition 
\begin{equation}
\int\limits_{m^{2}+\nu>0}dmf(m)=0 \,.  \label{int-f-zero}
\end{equation}
In the case of the $A_{2}$ contribution to the inequality (\ref{FPD-Y-ineq}),
the condition (\ref{int-f-zero}) holds due to Eq.~(\ref{c-constraint}).

\section{Conclusions}

In this paper, it is shown that the representation (\ref{H-mixture}) for the
quark-in-pion GPDs automatically obeys both polynomiality and positivity
constraints. This construction is based on the integration of the triangle
graphs for Yukawa model over the masses of the three propagators.
It also contains the
piece $(1-x)H_{\phi^{3}}$ whose positivity and polynomiality properties are
inherited from the triangle graph of the $\phi^{3}$ model. The mass integration allows a wide class
of mass dependent
weights constrained only by the positivity condition (\ref{s-from-c}), by
the divergence-cancellation requirement (\ref{c-constraint}) and
by the meson stability condition (\ref{meson-stable-a}).

We also have the freedom of adding an arbitrary $D$ term without violating
positivity and polynomiality. The possibility to include the $D$ term is
very important. Indeed, integrating over the masses of triangle graphs one
can generate only thresholds in the $t$ channel whereas the $D$ term allows
us to produce single-particle poles in the $t$ channel.

This paper describes only the method of the construction of GPDs obeying
polynomiality and positivity constraints. One can go beyond the
$\phi^{3}$ and Yukawa models trying to find new ``perturbative bricks'' for
the construction of the solutions of the positivity and polynomiality
constraints. One should keep in mind that in contrast to the two-point
correlation functions for which we have K\"{a}llen-Lehmann representation,
the case of GPDs is more involved and there is no guarantee
that the true physical GPD can be represented as an integral of triangle
graphs over their masses even if we go beyond the Yukawa model, include triangle
graphs from other theories and make our best from the freedom
to add an arbitrary $D$ term.

On the other hand, our construction is parametrized by arbitrary [up to the
constraints (\ref{meson-stable-a}), (\ref{c-constraint})] functions $a_{k}(m_{1},m_{3},\lambda)$
depending on three variables, i.e. our parametrization has the same amount
of ``degrees of freedom'' as the GPD $H(x,\xi,t)$ which also depends on three
variables. This means that the set of the solutions of the positivity
and polynomiality constraints covered by the representation
(\ref{H-mixture}) is rather large.

The comparison of the triangle graph approach considered here with the
formal mathematical solution of the positivity and polynomiality constraints
suggested in Ref.~\cite{Pobylitsa-02-d} shows a number of similar features
but at the moment it is not clear how large the overlap between the two
representations is. As long as this issue is not clarified it makes sense to
work with the ``superposition'' of the two representations. Indeed, from the
practical point of view the variety of structures compatible with the
polynomiality and positivity is more important than the problem of the
unambiguous parametrization of GPDs.

\textbf{Acknowledgments.} I am grateful to Ya.I. Azimov, A.V.~Belitsky,
M.~Diehl, L. Frankfurt, D.S.~Hwang, R. Jakob, P. Kroll, D. M\"{u}ller,
M.V.~Polyakov, A.V.~Radyushkin and M.~Strikman for useful discussions.
This work was supported by DFG and BMBF.

\appendix                                 

\section{Solution of the positivity bounds}

\label{r-k-appendix}

In this appendix we describe the properties of the variables $r_1,r_2$ and
derive the solution (\ref{pos-representation}) of the positivity
bound (\ref{ineq-q}).

The variables $r_{1},r_{2}$, which can be used instead of $x,\xi$, are
defined as follows:
\begin{equation}
r_{1}=\frac{1-x}{1+\xi },\quad r_{2}=\frac{1-x}{1-\xi }\,,  \label{r-12-def}
\end{equation}
\begin{equation}
\xi =\frac{r_{2}-r_{1}}{r_{2}+r_{1}}\,,\quad x=1-\frac{2r_{1}r_{2}}{
r_{1}+r_{2}}\,,
\end{equation}
\begin{equation}
\frac{2dxd\xi }{(1-x)^{3}}=\frac{dr_{1}dr_{2}}{r_{1}^{2}r_{2}^{2}}\,.
\label{u-v-measure}
\end{equation}
The region covered by the positivity bound (\ref{ineq-q})
\begin{equation}
x>|\xi |
\end{equation}
is mapped onto the square in the $r_{1},r_{2}$ plane
\begin{equation}
0<r_{1},r_{2}<1\,.
\end{equation}
Inequality (\ref{ineq-q}) takes the following form in terms of integration
variables $r_{1},r_{2}$ (we keep notation $x,\xi $ in GPDs implying that
these variables are functions of $r_{1},r_{2}$)
\[
\,\int_{0}^{1}\frac{dr_{1}}{r_{1}^{2}}\int_{0}^{1}\frac{dr_{2}
}{r_{2}^{2}}\left( \frac{r_{1}+r_{2}}{r_{1}r_{2}}\right) ^{N+1}p^{\ast
}\left( r_{2}\right) p\left( r_{1}\right) 
\]
\begin{equation}
\times \tilde{F}^{(N)}\left( x,\xi ,\frac{1-x}{1-\xi ^{2}}b^{\perp }\right)
\geq 0\,\,.  \label{uv-ineq-1}
\end{equation}
Since function $p$ is arbitrary we can replace it
\begin{equation}
p(r_{1})\rightarrow r_{1}^{N+3}p(r_{1})\,,
\end{equation}
which leads us to the equivalent form of inequality (\ref{uv-ineq-1})
\[
\int_{0}^{1}dr_{1}\int_{0}^{1}dr_{2}\left( r_{1}+r_{2}\right)
^{N+1}p^{\ast }\left( r_{2}\right) p\left( r_{1}\right) 
\]
\begin{equation}
\times \tilde{F}^{(N)}\left( x,\xi ,\frac{1-x}{1-\xi ^{2}}b^{\perp }\right)
\geq 0\,\,.  \label{uv-ineq-2}
\end{equation}
Inequality (\ref{uv-ineq-1}) means that the function
\begin{equation}
\left( \frac{r_{1}+r_{2}}{r_{1}r_{2}}\right) ^{N+1}\tilde{F}^{(N)}\left(
x,\xi ,\frac{1-x}{1-\xi ^{2}}b^{\perp }\right) 
\end{equation}
must be a positive definite quadratic form, i.e. it has the following
representation
\[
\left( \frac{r_{1}+r_{2}}{2r_{1}r_{2}}\right) ^{N+1}\tilde{F}^{(N)}\left(
x,\xi ,\frac{1-x}{1-\xi ^{2}}b^{\perp }\right) 
\]
\begin{equation}
=\sum\limits_{n}R_{n}(r_{1},b^{\perp })R_{n}^{\ast }(r_{2},b^{\perp })
\end{equation}
with some functions $R_{n}$. Turning back to the variables $x,\xi $ we find
\[
\tilde{F}^{(N)}\left( x,\xi ,b^{\perp }\right) =(1-x)^{N+1}
\]
\begin{equation}
\times \sum\limits_{n}R_{n}\left( \frac{1-x}{1+\xi },\frac{1-\xi ^{2}}{1-x}
b^{\perp }\right) R_{n}^{\ast }\left( \frac{1-x}{1-\xi },\frac{1-\xi ^{2}}{
1-x}b^{\perp }\right) \,.  \label{F-Q-deriv}
\end{equation}
In the case of real and $\xi $-even GPDs, the functions $R_{n}$ are real.
Introducing the functions 
\begin{equation}
Q_{n}(r,b^{\perp})=R_{n}\left( r,\frac{1}{r}b^{\perp}\right)\,,
\end{equation}
we obtain representation (\ref{pos-representation}) for $\tilde{F}
^{(N)}\left( x,\xi,b^{\perp}\right)$.

\section{Double distribution in the $\phi^3$ model}
\label{alpha-calculation}

This appendix contains a brief derivation of the double distribution
representation \cite{MRGDH-94,Radyushkin-96-a,Radyushkin-97}
for the GPD in the $\phi ^{3}$ model.
The contribution of the triangle graph of Fig.~\ref{fig-1}
with three different masses of partons is
\[
H_{\phi ^{3}}(x,\xi ,t|m_{1},m_{2},m_{3})\equiv (ig_{\phi ^{3}})^{2}
\]
\[
\times\int \frac{d^{4}q}{(2\pi )^{4}}\delta \left( x-1+
\frac{2q^{+}}{P_{1}^{+}+P_{2}^{+}}\right)
\]
\begin{equation}
\times 
\frac{i}{(P_{1}-q)^{2}-m_{1}^{2}+i0}
\frac{i}{(P_{2}-q)^{2}-m_{2}^{2}+i0}
\frac{i}{q^{2}-m_{3}^{2}+i0}
\,.
\label{H-phi-3-m123}
\end{equation}
The light-cone components $A^\pm$ of vectors $A^\mu$ are assumed to be chosen so that the
vector $n$ appearing in the definition of GPDs (\ref{F-N-def}) has only
one nonvanishing component $n^-$:
\begin{equation}
n^+ =0\,,\quad n^\perp=0\,.
\label{n-minus-only}
\end{equation}

Using the standard Feynman trick
\begin{widetext}
\begin{equation}
\prod\limits_{k=1}^{3}\frac{i}{q_{k}^{2}-m_{k}^{2}+i0}=\int_{0}^{1}d\alpha
_{1}\int_{0}^{1-\alpha _{1}}d\alpha _{2}
\int_{0}^{\infty }d\lambda
\,\lambda ^{2}
\exp \left\{ i\lambda \sum_{k=1}^{3}\alpha
_{k}(q_{k}^{2}-m_{k}^{2}+i0)\right\} _{\alpha _{3}=1-\alpha _{1}-\alpha _{2}}
\end{equation}
with
\begin{equation}
q_{1}=P_{1}-q\,,\quad q_{2}=P_{2}-q\,,\quad q_{3}=q\,,\quad
P_{1}^{2}=P_{2}^{2}=M^{2}\,,
\end{equation}
we find
\[
H_{\phi ^{3}}(x,\xi ,t|m_{1},m_{2},m_{3})=(ig_{\phi
^{3}})^{2}\int_{0}^{1}d\alpha _{1}\int_{0}^{1-\alpha _{1}}d\alpha
_{2}\int_{0}^{\infty }d\lambda \,\lambda ^{2}\int \frac{d^{4}q}{(2\pi
)^{4}}\delta \left( x-1+\frac{2q^{+}}{P_{1}^{+}+P_{2}^{+}}\right) 
\]
\begin{equation}
\times \exp \left( i\lambda \left\{ \alpha _{1}\left[
(P_{1}-q)^{2}-m_{1}^{2}\right] +\alpha _{2}\left[ (P_{2}-q)^{2}-m_{2}^{2}
\right] +(1-\alpha _{1}-\alpha _{2})\left( q^{2}-m_{3}^{2}\right) \right\}
\right)\,.
\label{H-phi3-double-calc}
\end{equation}
The calculation of the integrals over $q$ and $\lambda $ is straightforward
and yields
\[
\int_{0}^{\infty }d\lambda \,\lambda ^{2}\int \frac{d^{4}q}{(2\pi
)^{4}}\delta \left( x-1+\frac{2q^{+}}{P_{1}^{+}+P_{2}^{+}}\right) 
 \exp \left( i\lambda \left\{ \alpha _{1}\left[
(P_{1}-q)^{2}-m_{1}^{2}\right] +\alpha _{2}\left[ (P_{2}-q)^{2}-m_{2}^{2}
\right] 
\right.
\right.
\]
\[
\left.
\left.
+(1-\alpha _{1}-\alpha _{2})\left( q^{2}-m_{3}^{2}\right) \right\}
\right)
=\frac{1}{16\pi ^{2}}\delta \left( x-1+2\frac{\alpha _{1}P_{1}^{+}+\alpha
_{2}P_{2}^{+}}{P_{1}^{+}+P_{2}^{+}}\right) 
\]
\begin{equation}
\times \left\{ \alpha _{1}\alpha _{2}(P_{1}-P_{2})^{2}+(\alpha _{1}+\alpha
_{2})\left[ 1-(\alpha _{1}+\alpha _{2})\right] M^{2}-\left[ \alpha
_{1}m_{1}^{2}+\alpha _{2}m_{2}^{2}+(1-\alpha _{1}-\alpha _{2})m_{3}^{2}
\right] \right\} ^{-1}\,.
\label{q-lambda-int}
\end{equation}
Taking into account that according to Eqs.~(\ref{Delta-xi-t-def}),
(\ref{n-minus-only})
\begin{equation}
\frac{P_{1}^{+}}{P_{1}^{+}+P_{2}^{+}}=\frac{1+\xi }{2}\,,\quad \frac{
P_{2}^{+}}{P_{1}^{+}+P_{2}^{+}}=\frac{1-\xi }{2},\quad (P_{1}-P_{2})^{2}=t\,,
\end{equation}
we find from Eqs.~(\ref{H-phi3-double-calc}), (\ref{q-lambda-int})
\[
H_{\phi ^{3}}(x,\xi ,t|m_{1},m_{2},m_{3})=\frac{g_{\phi ^{3}}^{2}}{16\pi
^{2}}\int_{0}^{1}d\alpha _{1}\int_{0}^{1-\alpha _{1}}d\alpha _{2}\delta 
\left[ x-1+\alpha _{1}(1+\xi )+\alpha _{2}(1-\xi )\right] 
\]
\begin{equation}
\times \left\{
\left[ \alpha _{1}m_{1}^{2}+\alpha _{2}m_{2}^{2}+(1-\alpha _{1}-\alpha _{2})m_{3}^{2}\right]
-(\alpha _{1}+\alpha _{2})\left[1-(\alpha _{1}+\alpha _{2})\right] M^{2}
- \alpha _{1}\alpha _{2}t
\right\} ^{-1}
\,.
\label{H-phi3-m123-res}
\end{equation}

\end{widetext}

\section{Impact parameter representation for the GPD in the $\phi^3$ model}
\label{impact-triangle-appendix}

In this appendix we compute the triangle graph for the GPD of the $\phi^3$
model in the impact parameter representation in the region $x>|\xi|$.
In principle this could be done by applying the Fourier transformation
(\ref{F-impact-def}) to the double distribution representation
for this GPD (\ref{H-phi3-m123-res}).

However, we prefer another method based on the direct calculation of
the diagram in the impact parameter representation 
(see e.g. Ref.~\cite{Cheng-Wu-69}). The advantage of this approach
is that it explains the origin of the factorized form of the result.

We start from the Feynman integral (\ref{H-phi-3-m123}) for the triangle
graph of Fig.~\ref{fig-1}. One can integrate over $q^{-}$
deforming the integration contour. At $x>|\xi |$ the integral is determined
by the residue of the pole at $q^{2}=m_{3}^{2}$. Therefore at $x>|\xi |$ we
can replace on the RHS of Eq.~(\ref{H-phi-3-m123})
\begin{equation}
\frac{i}{q^{2}-m_{3}^{2}+i0}\rightarrow 2\pi \theta (q^{+})\delta
(q^{2}-m_{3}^{2})\,.
\end{equation}
\begin{widetext}
Then we have
\[
H_{\phi ^{3}}(x,\xi ,t|m_{1},m_{2},m_{3})
=(ig_{\phi ^{3}})^{2}\int \frac{d^{4}q}{(2\pi )^{4}}
\delta \left( x-1+\frac{2q^{+}}{P_{1}^{+}+P_{2}^{+}}\right) 
2\pi\theta (q^{+})\delta (q^{2}-m_{3}^{2})
\]
\begin{equation}
\times\frac{i}{(P_{1}-q)^{2}-m_{1}^{2}+i0}
\frac{i}{(P_{2}-q)^{2}-m_{2}^{2}+i0}
=\frac{g_{\phi ^{3}}^{2}}{2(1-x)}\int \frac{d^{2}q^{\perp }}{(2\pi )^{3}}\frac{
1}{(P_{1}-q)^{2}-m_{1}^{2}+i0}\frac{1}{(P_{2}-q)^{2}-m_{2}^{2}+i0}.
\label{H-impact-calc-1}
\end{equation}
On the RHS the components $q^{+}$ and $q^{-}$ are determined by the following equations
\begin{equation}
q^{+}=\frac{P_{1}^{+}+P_{2}^{+}}{2}(1-x),\quad q^{2}=m_{3}^{2}\,.
\end{equation}
It is straightforward to show that
\begin{equation}
(P_{k}-q)^{2}-m_{k}^{2}
=M^{2}+m_{3}^{2}-m_{k}^{2}-2(P_kq)
=M^{2}+m_{3}^{2}-m_{k}^{2}
-r_{k}\left| P_{k}^{\perp }-r_{k}^{-1}q^{\perp }\right| ^{2}-\left(
M^{2}r_{k}+m_{3}^{2}r_{k}^{-1}\right)
\label{poles-k} \,.
\end{equation}
Parameters $r_k$ are defined by Eq.~(\ref{r-12-def}). 
Using Eq.~(\ref{poles-k}), we can rewrite
Eq.~(\ref{H-impact-calc-1}) as follows:
\[
H_{\phi^{3}}(x,\xi ,t|m_{1},m_{2},m_{3})
=\frac{g_{\phi^3}^2}{2(1-x)r_1r_2}
\int \frac{d^{2}q^{\perp }}{(2\pi )^{3}}
\left[ \left| P_{1}^{\perp }-r_{1}^{-1}q^{\perp }\right|
^{2}+m_{3}^{2}r_{1}^{-1}(r_{1}^{-1}-1)+m_{1}^{2}r_{1}^{-1}-M^{2}(r_{1}^{-1}-1)
\right] ^{-1}
\]
\begin{equation}
\times \left[ \left| P_{2}^{\perp }-r_{2}^{-1}q^{\perp }\right|
^{2}+m_{3}^{2}(r_{2}^{-1}-1)r_{2}^{-1}+m_{2}^{2}r_{2}^{-1}-M^{2}(r_{2}^{-1}-1)
\right] ^{-1}\,.
\label{H-impact-calc-2}
\end{equation}
Now we define
\[
V\left( r_{k},b^{\perp }|m_k,m_{3}\right) =
\frac{g_{\phi^3}}{r_k}\int \frac{d^{2}p^{\perp }}{
(2\pi )^{2}}e^{ip^{\perp }b^{\perp }}\frac{1}{\left| p^{\perp }\right|
^{2}+m_{3}^{2}r_{k}^{-1}(r_{k}^{-1}-1)+m_{k}^{2}r_{k}^{-1}-M^{2}(r_{k}^{-1}-1)
}
\]
\begin{equation}
=\frac{g_{\phi^3}}{2\pi r_k}K_{0}\left( |b^{\perp }|r_{k}^{-1}
\sqrt{m_{3}^{2}(1-r_{k})+m_{k}^{2}r_k-M^{2}r_{k}(1-r_{k})}\right)\,,
\label{V-def-123}
\end{equation}
where $K_0$ is the modified Bessel function. Then it follows from
Eq.~(\ref{H-impact-calc-2})
\begin{equation}
H_{\phi ^{3}}(x,\xi ,t|m_{1},m_{2},m_{3})=\frac{r_{1}^2r_{2}^2}{4\pi(1-x)}
\int d^{2}b^{\perp }\exp \left[ ib^{\perp }\left( r_{1}P_{1}^{\perp
}-r_{2}P_{2}^{\perp }\right) \right] V\left( r_{1},r_{1}b^{\perp
}|m_{1},m_{3}\right) V\left( r_{2},r_{2}b^{\perp }|m_{2},m_{3}\right) \,.
\end{equation}
In the frame where $P_{1}^{\perp }+P_{2}^{\perp }=0$, we have
\begin{equation}
P_{1}^{\perp }=-P_{2}^{\perp }=-\frac{1}{2}\Delta ^{\perp }\,,
\qquad
r_1P_1^\perp-r_2P_2^\perp=-\frac{1-x}{1-\xi^2}\Delta^\perp
\end{equation}
so that
\begin{equation}
H_{\phi ^{3}}(x,\xi ,t|m_{1},m_{2},m_{3})=\frac{r_{1}^2r_{2}^2}{4\pi(1-x)}
 \int d^{2}b^{\perp }\exp \left[ -i\frac{1-x}{1-\xi ^{2}}\Delta
^{\perp }b^{\perp }\right] V\left( r_{1},r_{1}b^{\perp }|m_{1},m_{3}\right)
V\left( r_{2},r_{2}b^{\perp }|m_{2},m_{3}\right) \,.
\end{equation}
We see that in the impact parameter representation (\ref{F-impact-def})
our triangle graph for the GPD of the $\phi^3$ model has the following form:
\begin{equation}
\tilde{F}_{\phi ^{3}}\left( x,\xi ,\frac{1-x}{1-\xi ^{2}}b^{\perp
}|m_{1},m_{2},m_{3}\right) =\frac{1-x}{4\pi }V\left(
r_{1},r_{1}b^{\perp }|m_{1},m_{3}\right) V\left( r_{2},r_{2}b^{\perp
}|m_{2},m_{3}\right)\qquad (x>|\xi|) \,.
\label{tilde-F-res-m123}
\end{equation}

\section{Useful inequalities}

\label{inequalities-appendix}

In this appendix we derive two inequalities used in Sec.~\ref{FPD-section}.

{\em Inequality 1.} For any function $f$ and for any real constant $\nu$
\begin{equation}
\int\limits_{m_{1}^{2}+\nu >0}dm_{1}\int\limits_{m_{2}^{2}+\nu
>0}dm_{2}f(m_{1})f^*(m_2)
\frac{1}{m_{1}^{2}-m_{2}^{2}}\ln \frac{m_{1}^{2}+\nu }{m_{2}^{2}+\nu }
\geq 0
\end{equation}

{\em Proof. }Obviously
\[
\int\limits_{m_{1}^{2}+\nu>0}dm_{1}\int\limits_{m_{2}^{2}+
\nu>0}dm_{2}f(m_{1})f^*(m_2)\frac{1}{m_{1}^{2}-m_{2}^{2}}\ln \frac{
m_{1}^{2}+\nu}{m_{2}^{2}+\nu} 
=\int\limits_{m_{1}^{2}+\nu>0}dm_{1}\int\limits_{m_{2}^{2}+
\nu>0}dm_{2}f(m_{1})f^*(m_2)
\]
\begin{equation}
\times\int_{0}^{\infty }d\gamma \frac{1}{
(m_{1}^{2}+\gamma +\nu)(m_{2}^{2}+\gamma +\nu)} 
=\int_{0}^{\infty }d\gamma \left|\,
\int\limits_{m_{1}^{2}+\nu>0}dm_{1}f(m_{1})\frac{1}{m_{1}^{2}+\gamma +\nu}
\right| ^{2}\geq 0\,.
\end{equation}

{\em Inequality 2.} For any function $f$, obeying the condition 
\begin{equation}
\int\limits_{m_{1}^{2}+\nu>0}dm_{1}f(m_{1})=0  \label{int-f-zero-2}
\end{equation}
with some real $\nu$, we have
\begin{equation}
\int\limits_{m_{1}^{2}+\nu>0}dm_{1}f(m_{1})\int\limits_{m_{2}^{2}+
\nu>0}dm_{2}f^*(m_2)\frac{m_{1}^{2}+m_{2}^{2}+2\nu}{m_{1}^{2}-m_{2}^{2}}\ln 
\frac{m_{1}^{2}+\nu}{m_{2}^{2}+\nu}\leq 0\,.
\end{equation}

{\em Proof. }Omitting for brevity the integration region $m_{k}^{2}+\nu>0$,
we can write using Eq.~(\ref{int-f-zero-2})
\[
\int dm_{1}f(m_{1})\int dm_{2}f^*(m_2)\frac{m_{1}^{2}+m_{2}^{2}+2\nu}{
m_{1}^{2}-m_{2}^{2}}\ln \frac{m_{1}^{2}+\nu}{m_{2}^{2}+\nu} 
=2{\rm Re}\int dm_{1}f(m_{1})\int dm_{2}f^*(m_2)\frac{m_{1}^{2}+\nu}{
m_{1}^{2}-m_{2}^{2}}\ln \frac{m_{1}^{2}+\nu}{m_{2}^{2}+\nu} 
\]
\[
=2{\rm Re}\int_{0}^{\infty }d\gamma \left[ \int dm_{1}\frac{m_{1}^{2}+\nu}{
\gamma +m_{1}^{2}+\nu}f(m_{1})\right] \left[ \int dm_{2}\frac{1}{\gamma
+m_{2}^{2}+\nu}f^*(m_2)\right] 
\]
\[
=2{\rm Re}\int_{0}^{\infty }d\gamma \left[ -\int dm_{1}f(m_{1})+\int dm_{1}
\frac{m_{1}^{2}+\nu}{\gamma +m_{1}^{2}+\nu}f(m_{1})\right] 
\left[ \int dm_{2}\frac{1}{\gamma +m_{2}^{2}+\nu}f^*(m_2)\right] 
\]
\begin{equation}
=-2{\rm Re}\int_{0}^{\infty }d\gamma \left[ \int dm_{1}\frac{\gamma f(m_{1}) }{
\gamma +m_{1}^{2}+\nu}\right] \left[ \int dm_{2}\frac{f^*(m_2)}{\gamma
+m_{2}^{2}+\nu}\right] 
=-2\int_{0}^{\infty }d\gamma \,\gamma \left| \int dm_{1}\frac{f(m_{1})}
{\gamma +m_{1}^{2}+\nu}\right| ^{2}\leq 0\,.
\end{equation}
\end{widetext}

\end{document}